\def\year{2018}\relax
\documentclass[letterpaper]{article} %DO NOT CHANGE THIS
\usepackage{aaai18}  %Required
\usepackage{times}  %Required
\usepackage{helvet}  %Required
\usepackage{courier}  %Required
\usepackage{url}  %Required
\usepackage{graphicx,color}  %Required
\usepackage{listings}
\usepackage{amssymb}
\usepackage{amsmath}

\DeclareMathOperator{\var}{var}
\DeclareMathOperator{\cov}{cov}

\begin{document}

\title{A Modified L\'evy jump diffusion model Based on Market Sentiment Memory for Online Jump Prediction}

\author{Zheqing Zhu, Jian-guo Liu, Lei Li\\
Duke University\\
Durham, North Carolina 27708\\
}

\maketitle

\begin{abstract}
In this paper, we propose a modified L\'evy jump diffusion model with market sentiment memory for stock prices, where the market sentiment comes from data mining implementation using Tweets on Twitter. We take the market sentiment process, which has memory, as the signal of L\'evy jumps in the stock price. An online learning and optimization algorithm with the Unscented Kalman filter (UKF) is then proposed to learn the memory and to predict possible price jumps. Experiments show that the algorithm provides a relatively good performance in identifying asset return trends.
\end{abstract}

% A category with the (minimum) three required fields
%\category{H.4}{Information Systems Applications}{Miscellaneous}
%A category including the fourth, optional field follows...
%\category{D.2.8}{Software Engineering}{Metrics}[complexity measures, performance measures]

%\terms{Theory}

%\keywords{ACM proceedings, \LaTeX, text tagging} % NOT required for Proceedings

\section{Introduction}
Stock price is considered as one of the most attractive index that people like to predict \cite{steele12,PureJump}. An important model for the stock price that has been built in the academia is L\'evy process \cite{IntroLevy,ck03,ornthanalai2014}, which is a class of stochastic processes essentially show three features: a linear drift, a Brownian motion and a compound Poisson process. This model gains some success but it is pure random regarding the fluctuation.

With the external information of market and macroeconomics, a purely random compound Poisson process does not accurately reflect the fluctuation of the financial asset in a constantly changing financial environment.  The necessity of developing a model that incorporates external signal from the market is critical in improving accuracy in financial derivative pricing. One of the incentive for accurate pricing is to avoid financial crisis. In 2007-2008, part of the financial crisis was caused by unforeseen drop in option prices. Researchers have tried to develop distributions other than a normal distribution for pricing noise \cite{borland2002}, but not many have incorporated external signal for online learning and prediction.

 Instead of trying to develop a model which takes accurate noise into consideration, we aim to develop in this paper a modified L\'evy jump diffusion model with market sentiment memory to follow volatility clustering of financial assets, and a UKF algorithm to predict possible price jumps online. We intend to address the jump-diffusion effect in the financial market with big-data and machine learning technology to exploit market sentiment from Twitter. Different from previous approaches in financial asset pricing, the model involves non-Markovian processes with exponentially decaying memory, which can then be transformed into Markovian processes with higher dimension.

The main results of this paper include:

$\bullet$ Incorporating market sentiment memory which has memory in financial asset pricing and developing a modified L\'evy jump-diffusion model.
We take the market sentiment memory as the signal of L\'evy jumps in the pricing model (see Equations \eqref{eq:modifiedLevy} and \eqref{eq:Mdef}).

$\bullet$ Developing an unscented Kalman filter algorithm that actively learns market sentiment memory and accurately predicts asset trend accordingly (see Section {\it Market Sentiment Memory UKF Optimization Algorithm}).

$\bullet$ Capturing majority of big asset price movements (asset price jumps) from market sentiment with relatively high accuracy.
We found that outbreak of market sentiment indeed can predict majority of price jumps (see, for example, Figures \ref{fig:fb} (a) and (c)).

\section{Review of Current Work}
Much academic effort to model the stock market has been devoted to produce a better mathematical model based on Levy process and geometric Brownian model as foundations. Geman in 2002 used Levy process modeled with normal inverse Gaussian model, generalized hyperbolic distributions, variance gamma model and CGMY process, which reduces the complexity of underlying Levy measure, and produced meaningful statistical estimation of stock prices \cite{PureJump}. Cheridito in 2001 proposed fractional geometric Browniam motion model in order to gain a better estimate of stock price \cite{FBrown}.

Other academic efforts have been put in the field of sentiment analysis and prediction based on market data and public views. Sul et al. in 2014 collected posts about firms in S\&P 500 and analyzed their cumulative emotional valence and compared the return of firms with positive sentiment with other companies in S\&P 500 and found significant correlation \cite{TwitterTrading}. Bollen et al. in 2011 also produced similar result that Tweet sentiment and stock price are strongly correlated in short term \cite{TwitterMode}. Zhang and Skiena in 2010 used stock and media data to develop a automatic trading agent that was able to long and short stock based on sentiment analysis on media (Twitter and news platforms) data and getting a return as well as a high Sharpe ratio \cite{SentTrade}. Apart from pure sentiment analysis and trading strategies, Vincent and Armstrong in 2010 introduced prediction mechanisms also based on Twitter data to alert investors of breaking-point event, such as an upcoming recession \cite{TwitterBreak}.

In this paper, we propose a modified L\'evy jump diffusion model where we replace the standard compound Poisson component with a process determined by an exponentially decaying market sentiment memory, while the latter is extracted by UKF. UKF is able to take into account the non-linear transformation between market sentiment and asset return compared with the linear Kalman filter in \cite{duan1999estimating}, where Duan et al. proposed using linear Kalman filter to model exponential term structures in financial and economic systems.

\section{Background and preliminaries}
In this section, we give a brief introduction to the L\'evy jump-diffusion model for the price movement of assets and unscented Kalman filter. These are the building blocks for our modified L\'evy jump-diffusion model with market sentiment memory and its algorithm.
\subsection{L\'evy Jump-Diffusion Model}
In many theories such as Black-Scholes Model, the price movement of financial assets are modeled by the stochastic differential equation (SDE) driven by Brownian motion (\cite[Chap. 10]{steele12})
\[
dP(t)=\tilde{\mu}P(t)\,dt+\sigma P(t)\,dB(t) ,
\]
where $B(t)$ is a standard Brownian motion. The solution of this SDE is known to be the geometric Brownian motion
\[
P(t)= P(0)e^{\mu t+\sigma B_t},~~\mu=\tilde{\mu}-\frac{1}{2}\sigma^2.
\]
$\mu$ is called the drift factor of the geometric Brownian motion and  represents the log annual return of the financial asset. $\sigma$ represents the volatility of daily return of the asset. An interesting fact is that $\mu$ does not affect option pricing in the Black-Scholes model \cite{steele12,ross11}. However, if $\tilde{\mu}$ equals the interest rate $r$, this geometric Brownian motion becomes risk-neutral and gives the Black-Scholes formula for the no-arbitrage cost of a call option directly \cite[Chap. 7]{ross11}.

One of the drawbacks of the geometric Brownian motion is that the possibility of discontinuous price jumps is not allowed \cite{ck03,ornthanalai2014}. The L\'evy Jump-Diffusion Model is one that tries to resolve this issue (\cite[Chap. 8]{ross11}, \cite{ck03}):
\begin{equation}
    P(t) = P(0)e^{\mu t+\sigma B_t+\sum^{t}_{i = 1} (\sum_{j=1}^{N_i} J_{ij}-\lambda\kappa)},
\end{equation}
where $J_{ij}\sim N(\kappa, \sigma_J^2)$ i.i.d. are the jump parameters. $N_i\sim Poisson(\lambda)$ i.i.d. are the Poisson parameters that control the occurance of the jump. $\kappa, \sigma_J$ and $\kappa$ are constant parameters and $Z, N_i, J_{ij}$ are assumed to be mutually independent.
The log return for day $t$ is given by:
\begin{equation}\label{eq:loglevyjump}
    r(t) = \ln(P(t))-\ln(P(t-1)) = \mu + Z + (\sum_{j=1}^{N_t} J_{tj} -\lambda\kappa).
\end{equation}
The L\'evy jump-diffusion model consists of three components. The first part, $\mu$ is the expected logarithmic return of the financial asset. The second component, $Z$ is the white noise of the price or logarithmic return that is unpredictable. The third component, $\sum_{j=1}^{N_i} J_{ij} -\lambda\kappa$, is a compound Poisson distribution that provides a jump signal and a jump magnitude. The standard L\'evy jump diffusion model intends to include the fat tails that have been generally observed in asset returns in addition to the Gaussian structure that is commonly assumed. The downside of L\'evy jump diffusion model is that, although the model works well in a long-term structure, it fails to recognize market specific information that can be extracted from public opinion if used to predict short-term asset return.

\subsection{Unscented Kalman Filter}
In this section, we give a brief introduction to UKF (\cite{julier1997new,wanvan00}), which is used to give an accurate estimate of the state of nonlinear discrete dynamic system.
Suppose that the state of a system evolves according to
\begin{gather}\label{eq:dynamics}
x(t)=f_t(x(t-1), u(t))+b(t),
\end{gather}
where $x(t)$ represents the state of the system at time $t$, $u(t)$ is the external input and $b(t)$ is a noise with mean zero and covariance $Q_t$. $f_t$ is the known model of dynamics for $x(t)$. A measurement $z(t)$ is then made to $x(t)$:
\begin{gather}
z(t)=h(x(t))+d(t),
\end{gather}
where $d(t)$ is the measurement noise with mean zero and covariance $R_t$, independent of $b(t)$. $h$ is the known measurement function.

Suppose that $\hat{x}(t-1)$ is the belief of the state $x$ at time $t-1$ and the covariance matrix of $\hat{x}(t-1)$ is $P_{t-1}$.
The general process of Kalman filter is given as follows:
\begin{itemize}
\item Predict: Using the belief $\hat{x}(t-1)$, $P_{t-1}$, we obtain a predict (prior belief) of the state $\bar{x}(t)$. Denote $\bar{P}_t$ the covariance matrix of $x(t)-\bar{x}(t)$ (current prior belief in process covariance matrix).

\item Update: Using the prediction $\bar{x}(t)$, we have the predicted measurement $\mu_z$. When we have the observation $z(t)$, we update the belief of the state at $t$ as
\[
\hat{x}(t)=\bar{x}(t)+K_t(z(t)-\mu_z).
\]
$K_t$ is the Kalman gain matrix computed as follows: first of all, we compute the approximation of the covariance matrix of the residue $z(t)-\mu_z$ as $P_{z}$, and the covariance matrix between $x(t)-\bar{x}(t)$ and $z(t)-\mu_z$ as $P_{x z}$. Then, the Kalman gain matrix is given as
\[
K_t=P_{x z}P_{z}^{-1} .
\]
The intuition is that this is the ration between belief in state and belief in measurement.
The covariance matrix of $\hat{x}(t)$ is then computed as
\[
P_t=\bar{P}_t-K_tP_{z}K_t^{\top}.
\]
\end{itemize}

In the case $f_t$ and $h$ are nonlinear, it is usually hard to compute the mean and covariance of $\bar{x}(t)$ and $\hat{x}(t)$. The unscented transform computes these statistics as following: (i) generates $2n+1$ ($n\in\mathbb{N}_+$) (deterministic) sigma points using $\hat{x}(t-1)$, $P_{t-1}$ and $Q_t$, with certain weights $w_i^m$ and $w_i^c$. (ii) Evolve these sigma points under $f_t$, and obtain $2n+1$ data $Y_i$. Then, the statistics are approximated using these $2n+1$ data.
The UKF makes use of the unscented transform to approximate $\bar{x}(t)$, $\bar{P}_t$, $P_z$ and $P_{xz}$ to second order accuracy. Hence, the cycle of UKF can be summarized as following:

1. Predict the next state from the posterior belief in the last step: (UKF.predict())
\begin{equation}
\begin{split}
     X&=\Sigma(\hat{x}, P),\\ %  X &= sigma-function(x, P)\\
    Y_i &=  f_t(X_i, u(t)),~i=0,\ldots, 2n, \\
    \bar{x} & = \sum_{i = 0}^{2n} w^m_i Y_i\\
    \bar{P} & = \sum_{i = 0}^{2n} w^c_i(Y_i - \bar{x})(Y_i - \bar{x})^{\top} + Q,\\
\end{split}
\end{equation}
Here, $\Sigma$ is the algorithm to generate sigma points from the posterior belief $\hat{x}(t-1)$. The standard algorithm is Van der Merwe's scaled sigma point algorithm \cite{van2004sigma}, which, with a small number of sampling with corresponding weights $w_i^m$ and $w_i^c$, gives a good performance in belief state representation. See \cite{van2004sigma} for the formulas of $w_i^m$ and $w_i^c$.

2. Update from prior belief to posterior belief according to current noisy measurement. (UKF.update())
\begin{equation}
\begin{split}
    Z_i &= h(Y_i),~i=0, 1, \ldots, 2n, \\
    \mu_z &= \sum_{i=0}^{2n}w^m_iZ_i\\
    P_z &= \sum_{i=0}^{2n}w^c_i(Z-\mu_z)(Z-\mu_z)^{\top} + R_t \\
    K &= [\sum_{i=0}^{2n}w^c_i(Y-\bar{x})(Z-\mu_z)^{\top}]P_z^{-1}\\
    \hat{x} &= \bar{x} + K(z - \mu_z)\\
    P &= \bar{P} - KP_zK^T.\\
\end{split}
\end{equation}
Refer to \cite{julier1997new} for the details of implementation of a UKF. An important aspect of UKF that can be utilized in estimating jump-diffusion process is its Gaussian belief. We can regard each Gaussian component in the compound Poisson distribution in jump-diffusion as a state belief of UKF and apply to UKF to obtain a stable transition function with current sentiment data as input.

\section{Market sentiment}

\subsection{VaderSentiment}\label{sec:vadersentiment}

We aim to extract information from Twitter for both idiosyncratic sentiment information and market sentiment information. Before we present the implementation of data mining and analytics, we introduce VaderSentiment \cite{hutto2014vader}. For a given sentence $s$, VaderSentiment produces,
\begin{equation}
    vader(s) = [positive, neutral, negative, compound],
\end{equation}
where $positive$, $neutral$ and $negative$ are respective signals calculated and $compound$ is an overall evaluation of the quantitative sentiment of the sentence. Each of these signals are bounded within $[0, 1]$. Since we would like to extract instability of the sentiment analysis, we take vader's neutral as the noise and the confidence level is defined by the value of $1-neutral$. For each input sentence $s$, we extract:
\begin{equation}
\begin{split}
    sentiment(s) &= vader(s).compound\\
    noise(s) &= vader(s).neutral\\
\end{split}
\end{equation}
For a single day $t$, we have,
\begin{equation}\label{eq:conflevel}
\begin{split}
    S(t) &= \sum_{s\in Set(t)} (1-noise(s))*sentiment(s) ,\\
    E(t) &= \frac{\sum_{s \in Set(t)} noise(s)}{|Set(t)|} .\\
\end{split}
\end{equation}

\begin{figure*}
\centering
\includegraphics[width=0.9\textwidth]{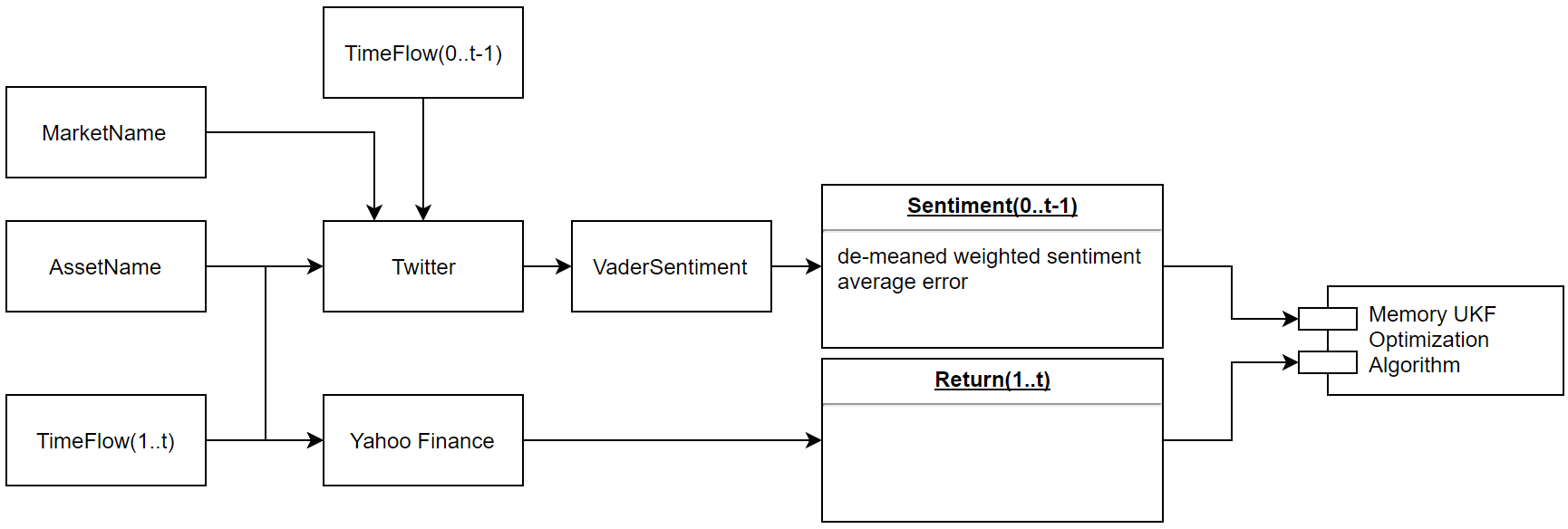}
\caption{Flowchart for Return-Sentiment Memory Kernel learning system.}
\end{figure*}

See Figure 1. At day t, from Twitter, we extract daily tweets from two pairs of keyword inputs, asset name (e.g. MSFT for Microsoft) and its trading market (e.g. NASDAQ) for idiosyncratic comments and trading sector (e.g. tech for Microsoft) and its trading market for market-related comments. For each day, we calculate idiosyncratic sentiment $S_I(t)$, macroeconomic sentiment $S_M(t)$, and the corresponding $E_I(t)$ and $E_M(t)$ from day t's Tweets using \eqref{eq:conflevel}. We then feed them into UKF sentiment memory optimization algorithm (which will be introduced in Section \ref{sec:modelalg} ) to learn the exponentially decaying memory model and then predict.

\subsection{Market Sentiment Memory}\label{subsec:marketsenti}

We now consider adding the memory process $m(t)$ of a random variable $\xi(t)$ into the model:
\begin{gather}
m(t)=\int_{-\infty}^t \gamma(t-s)\xi(s)\,ds+\int_{-\infty}^t \gamma_1(t-s)dB(s),
\end{gather}
where $\gamma$ and $\gamma_1$ are memory kernels and the second term represents the white noise. In this work, we assume throughout that the effect of white noise is negligible so that $\gamma_1=0$.

 In general, the kernel could be completely monotone.
By the Bernstein theorem, any completely monotone function is the superposition of exponentials \cite{widder41}. For approximation, we can consider finite of them:
\begin{gather}
\gamma(t)=\sum_{i=1}^N a_i \exp(-\lambda_i t).
\end{gather}
The advantage of these exponential kernels is that we can decompose $m$ as
\begin{gather*}
m(t)=\int_{-\infty}^t (\sum_{i=1}^N a_i \exp(-\lambda_i (t-s)) )\xi(s)\,ds \Rightarrow m=\sum_{i=1}^N m_i,
\end{gather*}
so that each $m_i$ satisfies the following SDE driven by $\xi$:
\begin{gather}\label{eq:singleSde}
dm_i=-\lambda_i m_i\,dt+a_i\xi(t)\,dt.
\end{gather}
In this way, the non-Markovian memory process is embedded into Markovian processes with higher dimension.

In this work, for simplicity, we just assume $N=1$ (i.e. the memory kernel is a single exponential mode) and consider the memories (denoted as $\eta_I, \eta_M$) of two individual sentiment inputs, idiosyncratic sentiment ($S_I(t)$) and macroeconomic sentiment ($S_M(t)$) so that the two sentiment processes are given by the discretized SDE \eqref{eq:singleSde}:
\begin{equation}\label{eq:memoryprocess}
\begin{split}
    \eta_I(t) &= p_I\eta_I(t-1) + a_IS_I(t),\\
    \eta_M(t) &= p_M\eta_M(t-1) + a_MS_M(t),\\
\end{split}
\end{equation}
where  $p\in [0, 1)$ is called unit decay factor, and $a>0$ is called the inclusion factor.
We define the market sentiment memory process $\eta(t)$ as a linear combination of two components $\eta_I$ and $\eta_M$ with
\begin{equation}\label{eq:eta}
    \eta(t) = c_I\eta_I(t) + c_M\eta_M(t).
\end{equation}
For algorithmic development purpose, we impose
\begin{gather}\label{eq:apconstraint}
p+a = 1,
\end{gather}
to limit the search space. Note that \eqref{eq:apconstraint} is not a constraint because we later we care $\kappa\eta$ only. Enforcing $p+a=1$ only selects a scaling for $\kappa$.

\section{Model and Algorithm}\label{sec:modelalg}
In this section we present a modified L\'evy jump model for the asset. A UKF is then used to predict jump magnitude on the next day using computed market sentiment.

\subsection{Modified L\'evy Jump Diffusion Model}
Recall the logarithmic return in L\'evy jump diffusion \eqref{eq:loglevyjump}. As we state in Section \ref{sec:vadersentiment}, we would like to incorporate external information from social media to extract market sentiment, which contributes to asset return movements. Here we define the modified L\'evy jump diffusion model,
\begin{equation}\label{eq:modifiedLevy}
    r(t) = \ln(P(t))-\ln(P(t-1)) = \mu + Z +(M -\nu),
\end{equation}
where $M$ is the jump amplitude random variable. $\nu$ is a constant to take off the drift trend in $M(t)$ (equivalent to $\lambda\kappa$ in the L\'evy jump diffusion model) and we compute $\nu$ in advance using history data.

We assume that the jump magnitude is determined by the total memory effect of market sentiment, or the market sentiment process $\eta$ in \eqref{eq:eta}. In particular, we assume:
\begin{equation}\label{eq:Mdef}
    M(t)=\kappa(t) \eta(t).
\end{equation}
which indicates that a current jump magnitude is determined by the sentiment memory. An implication of this setting is that market sentiment value from an individual day is a kind of volatile velocity to the return of an asset. We assume that $\kappa$ evolves with momentum  so that it satisfies the order 1 autocorrelation model (AR(1)):
\begin{gather}\label{eq:ar1}
\kappa(t)=\phi \kappa(t-1)+g+\epsilon_t,
\end{gather}
with $g$ being a constant for the innovation and $\epsilon_t$ being a discrete white noise.

\subsection{Market Sentiment Memory UKF Optimization Algorithm}\label{sec:ukfopt}

In this section, we introduce a UKF optimization algorithm. In the algorithm, the drift $\mu$ is determined in advance, which is the daily return in the long history, we set $g=1$ in \eqref{eq:ar1}, and preset $c$'s in Equation \eqref{eq:eta} for each iteration. The algorithm is used to find the optimal $p$,  $a$ and $\phi$ defined in Equation \eqref{eq:memoryprocess} and \eqref{eq:ar1} with in-sample data .

State of the system is represented as
\[
x(t)=\left[
\begin{array}{c}
r(t)\\
\kappa(t)\\
 \eta(t)\\
 \eta_I(t)\\
   \eta_M(t)
\end{array}
\right],
\]
and the input vector is
\[
u(t)=\left[
\begin{array}{c}
S_I(t)\\
S_M(t)
\end{array}
\right] .
\]
The dynamics of the system $f_t$ (see \eqref{eq:dynamics}) is given by:
\begin{equation}
f( x(t),  u(t+1); \Lambda)
 =
 \left[ \begin{array}{c}
 \mu + Z+ \kappa(t)\eta(t) -\nu\\
 \phi\kappa(t)+g\\
  c_I\eta_I(t) + c_M\eta_M(t)\\
  p_I\eta_I(t) + a_IS_I(t+1)\\
   p_M\eta_M(t)+a_MS_M(t+1)
\end{array} \right],
\end{equation}
with $\Lambda=[\phi, a_I, a_M, p_I, p_M]$ being the parameters.

We define true return on day $r_{*}(t)$ and
\[
\kappa_{*}(t)=(r_{*}(t)-\mu+\nu)/\eta(t)
\]
 as the measurements of $r(t)$ and $\kappa(t)$:
\begin{equation}
    h(x) = [r_{*}(t), \kappa_{*}(t)].
\end{equation}
Motivated by the fact that $S_I,S_M$ are random, we assume the measurement noise variance $R$ is a combination of that for $S_I$ and $S_M$:
\[
R = a_I^2 E_I^2(t) + a_M^2 E_M^2(t),
\]
 where $E_I$ and $E_M$ are confidence level of sentiment values computed by the second equation in \eqref{eq:conflevel}. The randomness in $S_I$ and $S_M$ provides noise for the evolution $f$ and $h$.

We introduce here Jenson's alpha and beta market risk to set $c_I$ and $c_M$ in \eqref{eq:eta}.
Beta market risk is defined as \cite{jensen1968performance}:
\begin{equation}
    \beta(t) = \frac{\cov(r(t), r_M(t))}{\var(r_M(t))},
\end{equation}
where $r_M$ is the market log return. Jenson's alpha is:
\begin{equation}
    \alpha(t) = r_i(t) - [r_f(t) + \beta(r_M(t)-r_f(t))] ,
\end{equation}
where $r_i$ is individual return and $r_f$ is risk free rate ($r_M$, $r_i, r_f$ are all computable from current data). Using Jenson's $\alpha$ risk, we set
\begin{equation}
\begin{split}
    c_I(t) &= \alpha(t)/r(t),\\
    c_M(t) &= 1 - c_I,\\
\end{split}
\end{equation}
for computing $\eta(t)$.

We define the objective function in the optimization.
\begin{equation}\label{eq:objective}
\begin{split}
    &U(JI_{UKF}, JI_{act})\\ =& \frac{|JI^{pos}_{UKF} \bigcap JI^{pos}_{act}| + |JI^{neg}_{UKF} \bigcap JI^{neg}_{act}|}{T} \\
    &- \frac{|JI^{pos}_{UKF} \backslash JI^{pos}_{act}| + |JI^{neg}_{UKF} \backslash JI^{neg}_{act}|}{T}.
\end{split}
\end{equation}

In the formula,
\[
JI=\{J_t: |r(t)-\mu|>1.96\sigma \}
\]
represents the set of jumps 1.96 standard deviations away from the process mean, or 1.96 volatility from the drift factor. $JI^{pos}$ indicates positive jumps and $JI^{neg}$ indicates negative jumps.
 What we aim to achieve here is that jumps identified by UKF overlaps the most with the actual jumps that happens. The major goal of UKF Optimization is essentially trying to identify a trend in the asset return time series.

With the settings above, we present the UKF optimization algorithm for searching optimal $p$, $a$ and $\phi$, with restriction that $p+a = 1$ and $\phi\in (0, 1)$.

\begin{lstlisting}
UKF_Optimize(coef_err, idio_sent[],
    mark_sent[], ret[]):
    Initialize x, P, Q, R
    p_I, p_M = 0
    Optimal = [a, p, 0]
    for p_I in 0..1 step coef_err:
       for p_M in 0..1 step coef_err:
         for phi in 0..1 step coef_err:
            for t in len(ret[]):
               UKF.predict(x, P,
                 f(p_I,p_M, phi))
               UKF.R = [(1-p_I)^2*
                   error_I^2(S_I(t))
                    + (1-p_M)^2*
                    error_M^2(S_M(t))]
               UKF.update(R(t+1))
            u = U(JI_UKF, JI_act)
            UpdateOptimal(a, p, phi, u)
    return Optimal.a, Optimal.p, Optimal.phi
\end{lstlisting}
where $U$ is the objective function of UKF-Optimization algorithm defined in \eqref{eq:objective}.  $\mathrm{UpdateOptimal(a, p,\phi, u)}$ means if $\mathrm{u}$ is bigger than the old $\mathrm{u}$, we update $\mathrm{(a,p,\phi)}$ to the new parameter and keep the old values otherwise.

We first use $\mathrm{UKF\_Optimize}$ with in-sample data to find the optimal $p_I$, $p_M$ and $\phi$ for the maximum coverage on the actual jumps. After the optimal parameters are obtained, we use UKF predict the Stoke price using Model \eqref{eq:modifiedLevy} online.

The UKF is generally used for state transition learning where the transition rules and noises are relatively stable. One reason is that during a near-stationary process, state belief is generally strengthened such that state transition converges. The Kalman gain factor, due to a strong belief in state, with very small covariance, quickly approaches 0. Consequently, UKF has learned pattern of state transition and is only mildly adjusted by input.  In our case, the economic process has different trends in different time windows while the UKF is hardly used to model a non-stationary process. A critical idea of learning the non-stationary economic model using UKF in our model is that, we would not want to model observations from the market as a sensor with fixed volatility. The volatility clustering effect of asset returns can greatly impact the training result. Here we model the volatility clustering effect with sentiment error term. The significance of this algorithm is that with a small modification, UKF can be used to learn multiple exponentially decaying sentiment memory with guaranteed process covariance convergence performance even given a chaotic non-linear system\cite{feng2007convergence} with a quadratic time complexity over the standard UKF by searching the coefficient space with some acceptable coefficient error. Note that the output of UKF is not a strictly exponentially decaying memory due to its non-pre-deterministic Kalman gain parameter.

\section{Experiment}
We now present the experimental results for  Facebook (FB), Microsoft (MSFT) and Twitter (TWTR).

\begin{figure}
\includegraphics[scale = 0.4]{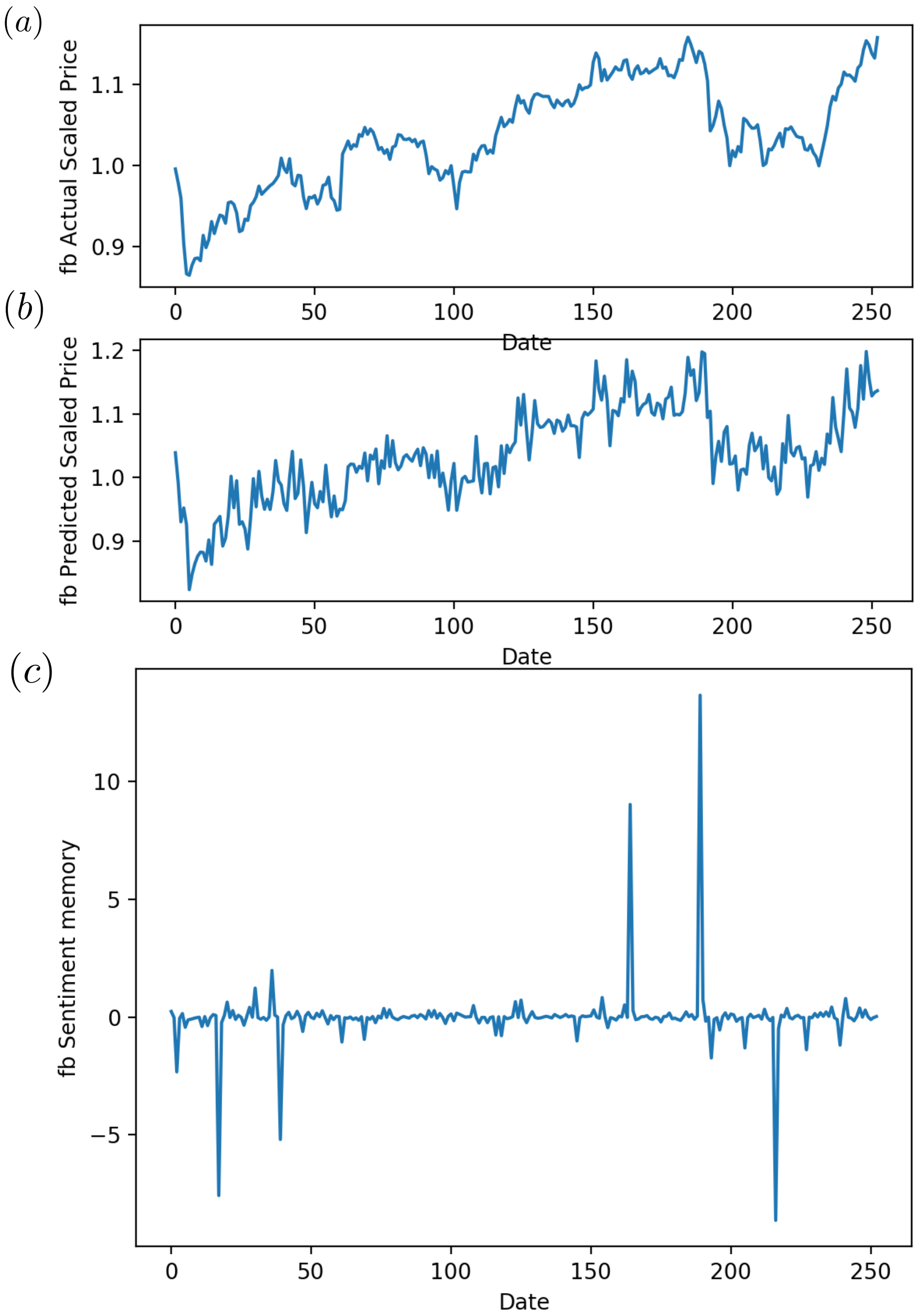}
\caption{Experimental results for FB}
\label{fig:fb}
\end{figure}

\begin{figure}
\includegraphics[scale = 0.4]{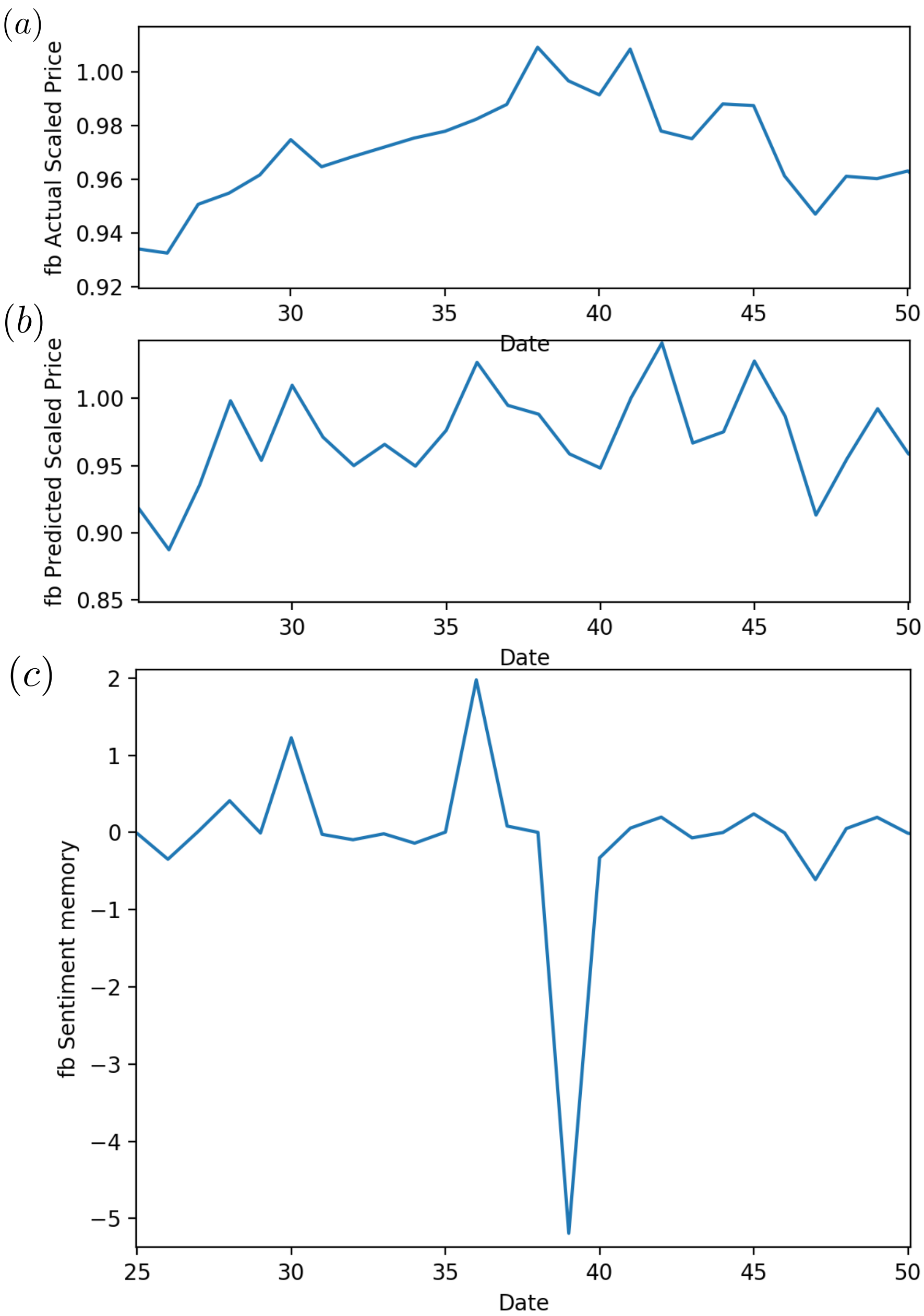}
\caption{Enlarged plots for FB}
\label{fig:fbenlarge}
\end{figure}

Figure \ref{fig:fb} (a) and \ref{fig:fb} (b) represent the actual return and the UKF return prediction based the modified L\'evy jump diffusion model for FB during the time period from 2016-02-03 to 2017-02-02  (note that there are only $252$ trading days in a year). The parameters $p, a, \phi$ are trained by the UKF optimization algorithm using date from 2013-02-02 to 2016-02-02. The jump prediction precision is $64.79\%$.   The in-sample prediction precision for time period 2013-02-02 to 2016-02-02 is $62.8\%$. Figure \ref{fig:fb} (c) shows $\eta(t-1)$ (we have offset $1$ in the memory plot because we use $\eta(t-1)$ to do the prediction for day $t$) .

The spikes in $\eta$ indicate outbreaks of market sentiment. To see how these spikes affect the jump prediction, we zoom in the plots for FB from Day $25$ to Day $50$ in Figure \ref{fig:fbenlarge}. There are evident spikes in $\eta(t-1)$ for $t=30, 36, 39, 47$. For $t=30$ and $t=47$, the real stock price curve has abnormal jumps, and our prediction of jumps based on the setiment memory has accurately predicted them. There is big outbreak of sentiment for $t=39$, and we can see that the real stock price goes down on Day $39$ and $40$. This indicates that the jumps in stock price curve are strongly correlated to market sentiment memory process and our model is able to predict a significant amount of abnormal jumps.

\begin{figure}
\includegraphics[scale = 0.4]{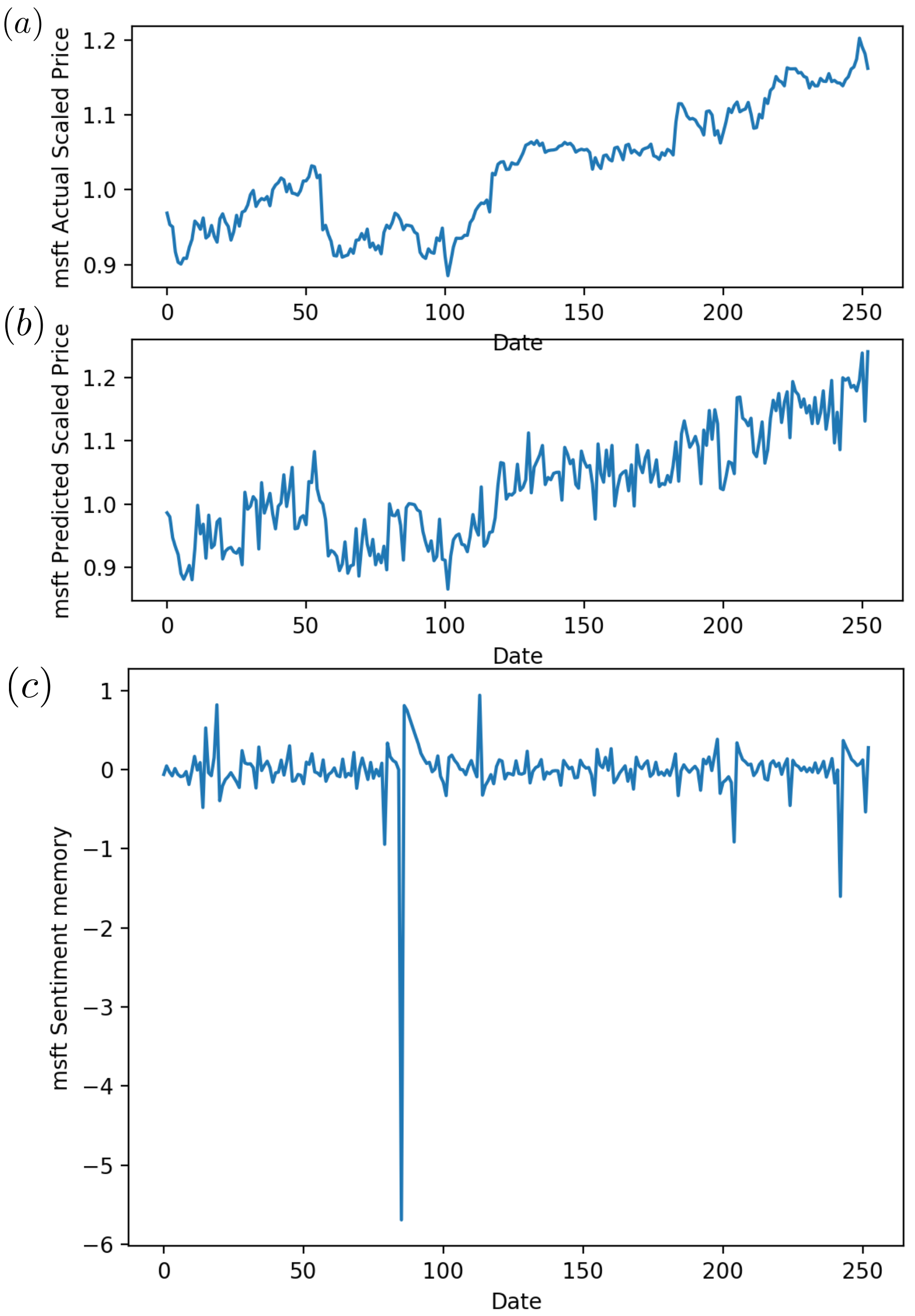}
\caption{Experimental results for MSFT}
\label{fig:msft}
\end{figure}

\iffalse
\begin{figure}
\includegraphics[scale = 0.4]{MSFT_enlarge.pdf}
\caption{Enlarged plots for MSFT}
\label{fig:msftenlarge}
\end{figure}
\fi

Figure \ref{fig:msft} (a) and \ref{fig:msft} (b) represent the actual return and the UKF return prediction for MSFT during the time period from 2016-02-02 to 2017-02-02, with training data from 2010-02-02 to 2016-02-02. The jump prediction precision is $52.96\%$ . The in-sample prediction precision for time period 2010-02-02 to 2016-02-02 is $64.2\%$. Figure \ref{fig:msft} (c) shows the market sentiment memory process $\eta(t-1)$ (Eq. \eqref{eq:eta}) .

\begin{figure}
\includegraphics[scale = 0.4]{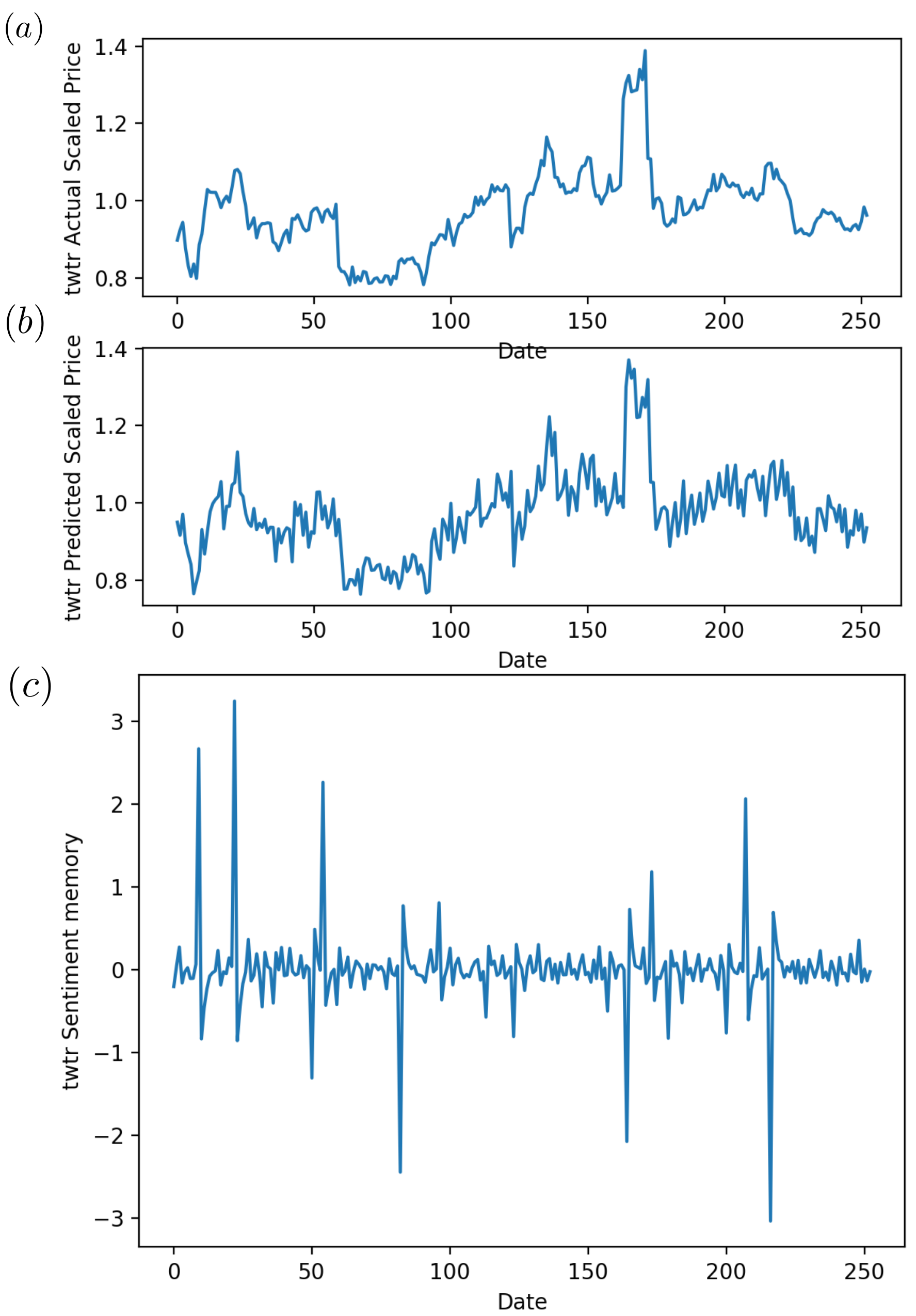}
\caption{Experimental results for TWTR}
\label{fig:twtr}
\end{figure}
Figure \ref{fig:twtr} (a) and \ref{fig:twtr} (b) represent the actual return and the UKF return prediction  for TWTR during the time period from 2016-02-03 to 2017-02-02, with the training data from 2014-02-02 to 2016-02-02.  The jump prediction precision is $60.85\%$.
The in-sample prediction precision for time period 2014-02-02 to 2016-02-02 is $65.4\%$. Figure \ref{fig:twtr} (c) shows the memory process $\eta(t-1)$.

There are a few significant observations we can draw from the results.

1. Using UKF generally captures the movement trend of the underlying asset, with little guidance with daily returns. More specifically,
during periods where sentiment memory kernel value peaks, the stock asset's return has very strong correspondence. However, when there is very minor value in market sentiment kernel, the asset return prediction follows the previous trading day's return, triggering some inaccuracy.

2. Movements in UKF return prediction are in general greater in magnitude than actual returns. This could be caused by high volatility of sentiment values.

3. From the sentiment memory graphs (Figures \ref{fig:msft}- \ref{fig:twtr} (c)), we can observe a strong indication of clustering, which is an evidence of a decaying memory, analogous to GARCH model \cite{bollerslev1986} which measures volatility clustering. This can also be confirmed by trained parameters from UKF-optimization algorithm:\\
MSFT: $p_I = 0.11, p_M = 0.87,  \phi=0.63$.\\
FB: $p_I = 0.55, p_M = 0.36,  \phi=0.41$.\\
TWTR: $p_I = 0.47, p_M = 0.58,  \phi=0.84$.\\

\section{Discussion}
In this paper, we propose a modified L\'evy jump diffusion model with market sentiment memory for stock prices.  An online learning and optimization algorithm with UKF is used to predict possible price jumps. The result from the experiments instantiate our theory in market sentiment memory and its impact on asset returns. Our work has significance in both economics and computer science.

Regarding economics, our experiments have shown the existence of predictability in return by sentiment, which indicates market inefficiency in digesting public sentiment. The impact of market sentiment memory on asset returns can dramatically change the pricing models for options and financial derivatives because currently most of these products rely on the Markovian assumption about financial assets. To incorporate market sentiment memory into the pricing models, one possible way is to multiplying previous jumps occurring in asset's return history with decaying factors and then add the models, since jumps are strong indicators of market sentiment outbreak. Another possible way is to include a time series of market sentiment with explicit values into asset pricing models. Clearly, our model adopts the second strategy.

Regarding computer science, our work indicates that Kalman filter techniques (especially UKF) allow online learning for non-observable variables. The market sentiment memory can not be measured directly and it is an indirect variable, however, unlike other machine learning techniques, UKF allows online learning of such indirect variables in an iterative manner.

\iffalse
\section{Conclusion}
In this paper, we present an active learning and optimization search algorithm with Unscented Kalman Filter to identify underlying market sentiment memory kernel. The algorithm with data mining implementation using Tweets on Twitter, provides a relatively good performance in identifying asset return trends.
\fi

\bibliographystyle{aaai}
\bibliography{main}

\end{document}